\RequirePackage{fix-cm}
\documentclass[smallextended]{svjour3}       
\smartqed  
\usepackage{graphicx}
\begin{document}

\title{The orbit of Aegaeon and the 7:6 Mimas-Aegaeon resonance
\thanks{NCJ is grateful to Fapesp (Sao Paulo State Research Funding Agency), for the processes 2019/15162-2, 2020/06807-7. AR is grateful to FAPERJ (process 210.419/2022).}
}


\author{Nelson Callegari Jr.         \and
        Adri\'an Rodr\'iguez 
}


\institute{Nelson Callegari Jr. \at
             S\~{a}o Paulo State University (Unesp), Institute of Geosciences and Exact Sciences (IGCE). Avenue 24-A, 1515, Rio Claro, SP; Zip code 13506-900, Brazil.\\
              Tel.: +55-19-35269132\\
              \email{nelson.callegari@unesp.br}           
           \and
            Adri\'an Rodr\'iguez \at
              Observat\'orio do Valongo, Universidade Federal do Rio de Janeiro. Ladeira do Pedro Ant\^{o}nio 43, Rio de Janeiro, RJ;  Zip code 20080-090, Brazil.\\
               \email{adrian@ov.ufrj.br}
               }

\date{Received: date / Accepted: date}

\maketitle

\begin{abstract}

Aegaeon (S/2008 S 1) is the last satellite discovered by the Cassini spacecraft at the end of the 2000s. Like the satellites Methone and Anthe, it is involved in mean motion resonance with the mid-sized Mimas.

In this work, we give a detailed analysis of the current orbit of Aegaeon identifying the resonant, secular and long-term perturbations due to Mimas and the oblateness of Saturn, and the effects of Tethys. For this task, we perform thousands of numerical simulations of full equations of motion of ensembles of small bodies representing clones of Aegaeon.

We have mapped the domain of the 7:6 Mimas-Aegaeon resonance in the phase space of the semi-major axis and eccentricity. It displays a large area dominated by regular motions associated with the 7:6 corotation resonance surrounded by chaotic layers. Aegaeon is currently located very close to the periodic orbit of the resonance, which extends up to eccentricities $\sim0.025$ centered at semi-major axis $\sim168,028$ km.

We show that the current orbit of Aegaeon has an important forced component in eccentricity due to the 7:6 resonance. The orbital inclination of Aegaeon has a non-negligible forced value due to long-term perturbations of Mimas. These two forced modes explain the complex perturbed orbit of Aegaeon without requiring the co-existence of multiple resonances.

\end{abstract}

\section{Introduction}

Throughout the 2000s the \emph{Cassini} space probe discovered the new close-in satellites Methone (S/2004 S 1; Porco 2004), Anthe (S/2007 S 4; Porco 2007), Aegaeon (S/2008 S 1; Porco 2009), and rediscovered Pallene (S/2004 S 2; Porco 2004), which had already been detected in \emph{Voyager}'s images (see Spitale et al. 2006). After that, studies on these intriguing small objects (with diameters less than 10 km) relied on their origins (e.g. Porco et al. 2007, Charnoz et al. 2010), ring-satellite interactions and stability of particles (e.g. Hedman et al. 2009; Madeira et al. 2018, Rodr\'{i}guez and Callegari 2021); surfaces characteristics (e.g. Hedman et al. 2010, Thomas and Helfenstein 2020); or resonant dynamics, phase space properties and long-term stability (Hedman et al. 2010, El Moutamid et al. 2014; Mun\~oz-Guti\'errez and Giuliatti-Winter 2017; Callegari and Yokoyama 2010a, 2020; Callegari et al. 2021; Rodr\'{i}guez et al. 2022, submitted - this Issue).

In this work, we aim to continue with the latter branch of research listed above by studying in detail the orbital dynamics of Aegaeon. Our primary motivation is to give continuity to our investigations of the so-called ``corotation resonances'', which correspond to that dynamical regime where the critical angle associated to resonance librates around the longitude of the pericenter (or of the apocenter) of the disturbing body. In the Saturnian system, Anthe, Methone, and Aegaeon are currently trapped in the domains of the 11:10 Anthe-Mimas, 15:14 Methone-Mimas, and 7:6 Mimas-Aegaeon mean motion corotation resonances, respectively. Mimas is the perturbing body since it is a mid-sized satellite with almost 400 km in diameter, representing the dominant mass in the above resonant pairs. In the case of the 7:6 resonance, the critical angle $\sigma_2=7\lambda_M-6\lambda_{Ae}-\varpi_M$ librates around $\pi$ with relatively small amplitude (Hedman et al. 2010), where the subscripts $M$, $Ae$ refer to Mimas and Aegaeon, respectively; $\lambda$ and $\varpi$ are the mean longitude and the longitude of the pericenter.

Numerical mappings of the whole domains of the 11:10 Anthe-Mimas and 15:14 Methone-Mimas resonances have been studied in detail by Callegari and Yokoyama (2020) and Callegari et al. (2021), respectively. In this work, we will focus entirely on the 7:6 Mimas-Aegaeon resonance. It is worth emphasizing that the three resonances have been studied by El Moutamid et al. (2014) by adopting average models and numerical techniques like surfaces of section. Mun\~oz-Guti\'errez and Giuliatti-Winter (2017) and Rodr\'{i}guez and Callegari (2021) also explored the phase space of the three resonances, but their approaches involve the dynamical lifetime of assemblies of particles surrounding Anthe, Methone and Aegaeon.

The main results on the dynamics of Aegaeon are given in Section 2. A global view of the Aegaeon phase space obtained with dynamical maps is introduced in Section 2.1. The mapping of the resonance has been done numerically with a model which includes several satellites disturbing the Aegaeon's orbit, being more general than previous studies of the phase space of the 7:6 resonance (e.g. El Moutamid et al. 2014). A close view of the current orbit of Aegaeon is given in Section 2.2. In particular, we show that the current resonant state is very close to equilibrium solutions associated to the corotation resonance, explaining the small amplitude ($\sim15$ degree) of libration of $\sigma_2$. In Sections 2.2 and 2.3, we isolate the main gravitational perturbations of the orbit of Aegaeon through numerical Fourier analysis. They are physically interpreted in terms of corotation resonance and forced resonant components in eccentricity and inclination.

Hedman et al. (2010) reported seven other critical arguments of interest in the disturbing function associated with the 7:6 resonance, besides the corotation angle. They refer these angles by $\varphi_{ILR}$, $\varphi_x$, $\varphi_y$, $\varphi_a$, $\varphi_b$, $\varphi_c$, $\varphi_d$. (See Table 1 in Section 2.4 for definitions. In this work, we utilize a different notation ($\sigma_i$) for the critical angles.) Adopting a different methodology, we also isolated \emph{the same} set of critical arguments with physical significance selected by Hedman et al. (2010). Although neither of these angles permanently librate, Hedman et al. (2010) explain their intriguing time variations in terms of multiple resonances acting on the orbit of Aegaeon. We follow the steps of the two previous works on Anthe and Methone (Callegari and Yokoyama 2020, Callegari et al. 2021), and interpret the perturbing components on the orbit of Aegaeon as a result of the composition of several fundamental frequencies of the system involving the corotation resonance and the long-term time variations of the Aegaeon and Mimas orbits. This discussion is given in Section 2.4.

Hedman et al. (2010), exploring the orbit of Aegaeon with initial values within the error uncertainty of the vector state of Aegaeon at that epoch, report that the close vicinity of its present orbit shows a rich dynamical environment in numerical simulations. By adopting a different methodology (we varied the initial orbital elements around the current ones at the epoch rather than vector states), we give evidence that the Aegaeon phase space at initial conditions located relatively far from the current neighborhood reveals other regions of physical interest. We denote these loci of the phase space by $\texttt{D}$ and \textbf{S} and orbits within them will be investigated in Section 3.1 and 3.2, respectively.

In Section 4, we close the paper with the main conclusions and some discussion.

\section{Dynamical Maps and the 7:6 Mimas-Aegaeon resonance}

\subsection{A global view of the 7:6 resonance}

As pointed out in Section 1, one of the main goals of this work is to study the global dynamics of the 7:6 Mimas-Aegaeon mean motion resonance. For this task, we utilize the dynamical mapping of the phase space. Figs. \ref{<F1>}a,b,c show three examples of dynamical maps. The initial conditions are taken in grids, and the orbits are numerically obtained with the numerical integrator RA15 (Everhart 1985). For each initial condition, the map is obtained after spectral analysis of a determined variable of the problem (denoted by VA), like semi-major axis, eccentricity, or inclination. For this task, we apply the Fast Fourier Transform method (e.g. Press et a. 1996).

The model adopted in numerical simulations consists of full equations of motion, considering Saturn, clones of the satellite Aegaeon, Mimas, Enceladus, Tethys, and Dione, and we take into account the non-central components $J_2$ and $J_4$ of the Saturn's gravitational field. The details on the equations of motion and reference system adopted in this work are given in Callegari and Yokoyama (2010b). To reduce computational time in the integrations of dense mappings, Rhea and Titan may not be included in general since their perturbations can be neglected in the sense that the main structures of the map are preserved in the absence of these satellites. The latter conclusion is based on a great deal on numerical experiments; however, the pair Rhea-Titan is considered when individual numerical integration devoted determining the main frequencies of the Aegaeon are performed since these cases do not require large computational efforts (e.g. Figs. \ref{<F4>}, {\ref{<F5>}). It is worth noting that only the initial conditions of the grid are changed, in general, the initial eccentricity versus the initial semi-major axis in the plane $(a_0,e_0)$, so that all other initial elements of the clones of Aegaeon and the other satellites are always fixed at the date January 01, 2016. The initial values have been taken from the \emph{Horizons} system of ephemerides, provided by JPL/NASA. See Appendix.

The spectra are obtained in such a way that, for each initial condition, we associate a spectral number, $N$, consisting of the number of peaks in that spectrum that are larger than a previously determined percentage of the highest peak. This percentage will be denoted by RA, the reference amplitude, and it is usually $1\%$ or $5\%$ of the largest peak. In the account of $N$ we exclude those peaks linked to short period smaller than 15 day so that the dynamical maps are free from short-term variations, and this may have important consequences in their interpretation. The dynamical maps are constructed in such a way that we associate a color palette where distinct tones are linked with different values of $N$. We must choose a cutoff value of $N$, say $N^*$, which is arbitrary in general, but must be chosen so that when $N>N^*$, the same color is considered. Additional details on this technique can be seen in Callegari et al. (2021).

 \begin{figure}
 \centering
 \includegraphics[width=12cm]{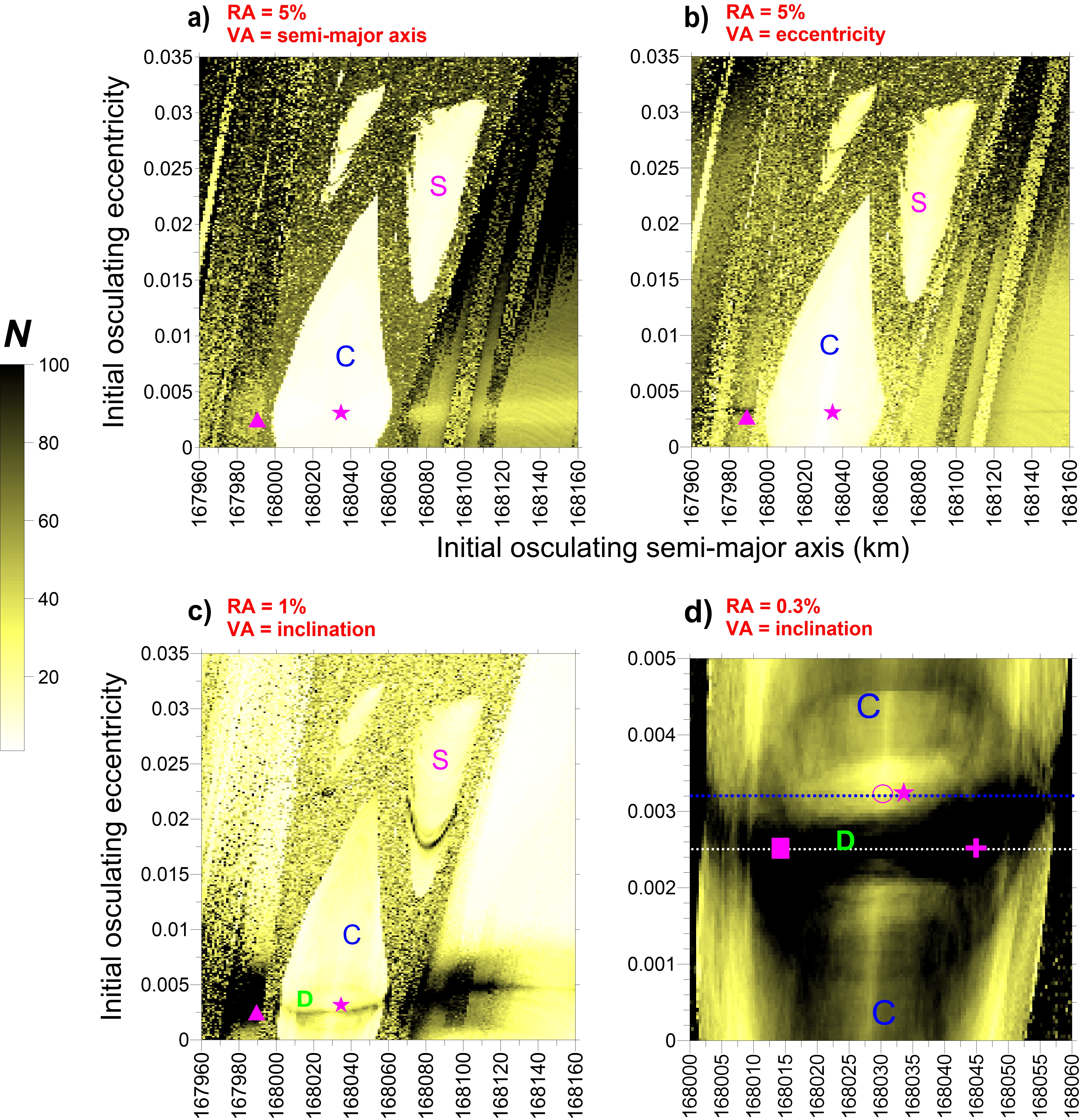}
  \caption{Dynamical mappings around the current orbit of Aegaeon at date January 01, 2016, where $a_0\sim168,033.2$ km, $e_0\sim0.0032$ (indicated by \textbf{a} magenta star). \textbf{C} is the Corotation zone studied in Sections 2.1 and 2.2. The black strip $\texttt{D}$ and the region \textbf{S} are analyzed in Sections 3.1 and 3.2, respectively. The integration time for each initial condition is 287.08 year, \textbf{a} time step of $0.1$ day.  $N$ is the spectral number defined in Section 2.1. In \textbf{(a,b,c)}: Maps with $\Delta a=1.25$ km, $\Delta e=0.00014$, and the reference amplitude RA utilized in the variable (VA) are \textbf{(a)} semi-major axis (RA=$5\%$), \textbf{(b)} eccentricity (RA=$5\%$), \textbf{(c)} inclination (RA=$1\%$). The triangle magenta symbol in \textbf{(a)} has initial conditions $(a_0,e_0)=(167,990,0.0025)$. \textbf{(d)}: Detailed at the bottom of \textbf{(c)}, where a total of 15,251 initial conditions have been integrated, and RA is $0.3\%$. Magenta symbols correspond to different $(a_0,e_0)$. Square: $(168,014,0.0025)$; crux $(168,045,0.0025)$; open circle $(168,030,0.0032)$. }
    \label{<F1>}
\end{figure}

The dynamical maps constructed in the way described above are very useful to detect regions of phase space with significant physical meaning. Several of these regions are shown in Fig. \ref{<F1>}. In  Figs. \ref{<F1>}a,b,c the ranges $\Delta a=200$ km in semi-major axis [167,960 km - 168,160 km] and $\Delta e=0.035$ in eccentricity have been adopted. The magenta stars in the maps show the location of Aegaeon on the adopted date (January 01, 2016). At this date, the initial osculating values of the semi-major axis, eccentricity, and inclination (w.r.t. Saturn's equator) are $a_0\sim168,033.2$ km, $e_0\sim0.0032$, $i_0\sim0.001$ degree. The maps in Figs. \ref{<F1>}a,b,c have been generated after numerical integration and Fourier analysis of 40,411 orbits.

The structure having the format of a candlelight in the vicinity of Aegaeon in the interval [168,000 km - 168,060 km] and reaching 0.023 in eccentricity, represents the domain of the Corotation zone associated with the 7:6 Mimas-Aegaeon mean motion resonance. In the \emph{whole} region, indicated by $\textbf{C}$ in Fig. \ref{<F1>}, the critical angle $\sigma_2=7\lambda_M-6\lambda_s-\varpi_M$ librates around $\pi$, where $s$ indicates a test satellite clone of Aegaeon.

The maps reveal that the interior of the Corotation zone is generally whitish, with small values of the spectral number $N$. This occurs since the motion is regular due to the trapping into resonance. An exception of this regular aspect occurs at the bottom of the Corotation zone when the inclination of the test satellites is Fourier analyzed (Fig. \ref{<F1>}c). In this case, a black horizontal strip can be seen for orbital eccentricities very close to the current one at the date. This structure is denoted by $\texttt{D}$ in the maps.

At larger eccentricities in the maps, other two regular regions appear isolated and separated from the Corotation zone by black regions associated with their separatrices. Note that in the inclination map, there is also a separatrix inside the regular region at the right indicated by \textbf{S}. The analysis of dynamics within this region will be devoted to Section 3.

Rodr\'{i}guez and Callegari (2021) consider the mapping of Aegaeon’s phase space adopting the technique of calculation of mean values of $\Delta a$, $\Delta e$, $\Delta i$, for each numerical simulation. The results of Fig. \ref{<F1>} can be compared with figure 5 in Rodr\'{i}guez and Callegari (2021). There is good accordance in the results of the determination of the boundaries of the 7:6 Mimas-Aegaeon resonance, but the fine structures are better determined with mapping in the frequency domain.

In the following (Sections 2.2, 2.3, and 2.4), the current orbit of Aegaeon will be fully explored.

 \begin{figure}
 \centering
 \includegraphics[width=12cm]{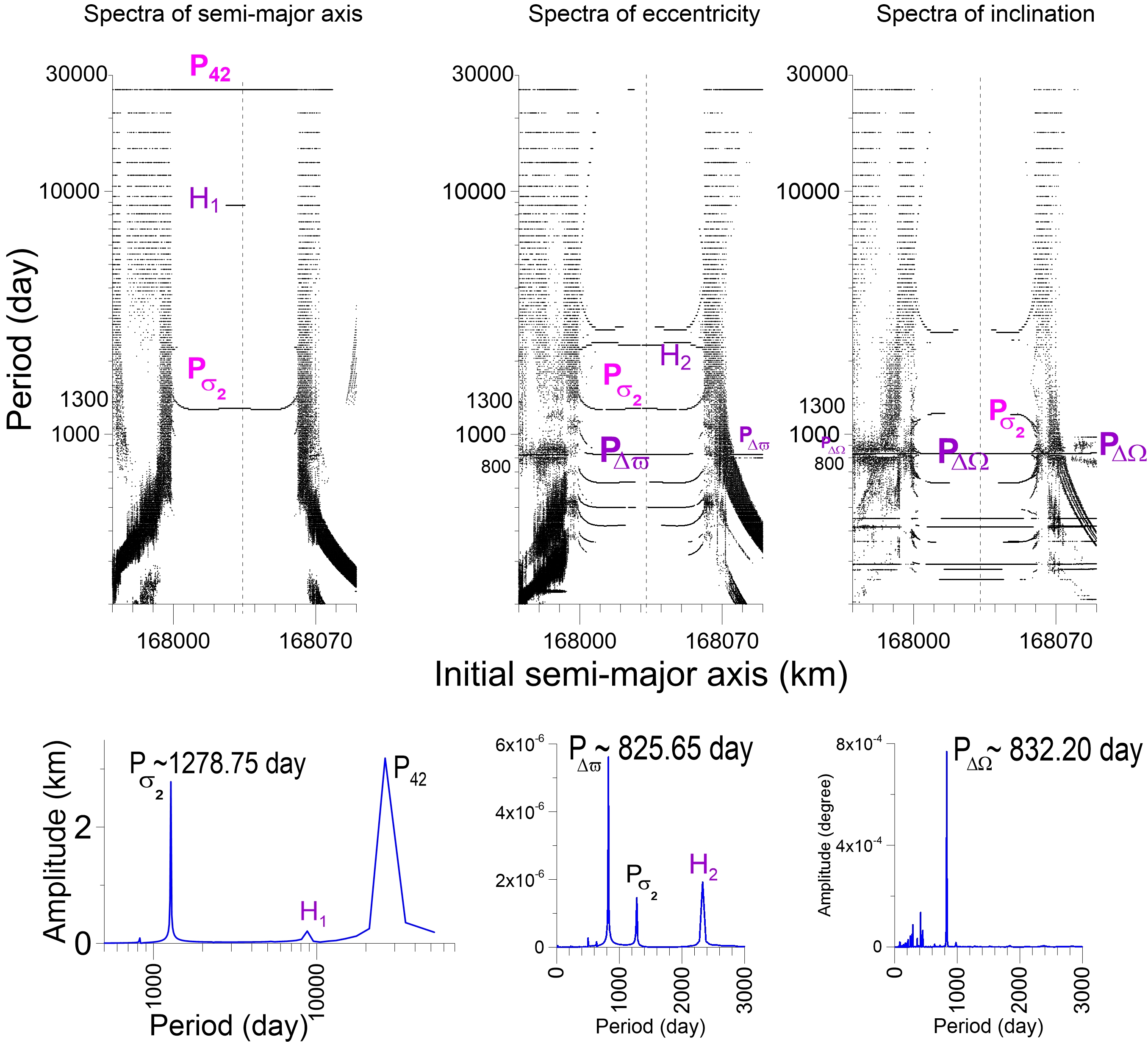}
  \caption{\textbf{Top row}: Individual Power Spectra (IPS) of 1500 clones of Aegaeon constructed from spectra of different variables indicated at the top of the plots. \textbf{$P_{\sigma_2}$} is the period corresponding to the 7:6 Mimas-Aegaeon Corotation resonance. \textbf{$P_{42}\sim26,214.4$} is associated to Mimas-Tethys 4:2 resonance. \textbf{$P_{\Delta\varpi}$}, \textbf{$P_{\Delta\Omega}$}: Mutual-secular long-term Mimas-Aegaeon perturbations. \textbf{$H_1$} and \textbf{$H_2$} are harmonics. \textbf{Bottom row}: Individual spectra of the orbit corresponding to the initial semi-major axis indicated by a vertical dashed line in the top row, similar to the current value of Aegaeon at January 01, 2016. Integration time is 1720 year, time step 0.12 day. The approximated values of the main periods are indicated close to the peaks. The x-axis of the spectrum of the semi-major axis is shown with a logarithmic scale.}
  \label{<F3>}
\end{figure}

\subsection{The current orbit of Aegaeon: resonant perturbations and the Corotation zone.}

Fig. \ref{<F1>}d shows the bottom part of the Corotation zone in more detail, allowing us to identify more refined structures, like the loci of the equilibrium solution associated with the Corotation zone, \textbf{C}. It appears like an almost vertical yellow line, reaches $e_0=0$ close to $a=168,028$ km, and crosses the region $\texttt{D}$ at $e_0\sim0.0025$. We can note that the position of Aegaeon in the map is very close to the periodic orbit, a result that agrees with Hedman et al. (2010).

Our dynamical maps are constructed after analyses of the spectra of the orbits. We can have a more quantitative knowledge of the distribution of the frequencies in the phase space by making unidirectional sweepings of the resonance by adopting the initial semi-major axis of test satellites as a free parameter in numerical simulations. The results are the ``individual dynamical power spectra'', denoted by IPS. The y-axis of an IPS shows, for each initial condition, the periods associated with the peaks in the spectrum with amplitudes larger than a prefixed fraction (the RA amplitude defined before).

For this task, we must fix the initial eccentricity and the other initial elements. Let us begin with the current osculating eccentricity of Aegaeon on January 01, 2016, indicated by the blue dotted line $e_0=0.0032$ in Fig. \ref{<F1>}d. Fig. \ref{<F3>} shows three IPSs calculated from the spectra of the semi-major axis (top row), eccentricity (middle), and inclination (right) of 1500 test satellites in the interval $167,970\leq a_0\leq 168,090$ km. RA is $1\%$, and y-axes are given in logarithmic scale. The model and integration time are the same as defined in Fig. \ref{<F1>}. The loci of the fundamental periods are distributed in the vertical direction and identified with different symbols.

The isolated (almost) horizontal line in Fig. \ref{<F3>}, top-left, is the functional dependence of the period of libration of $\sigma_2$ with $a_0$, and it is indicated by $P_{\sigma_2}$. There is a smooth variation of $P_{\sigma_2}$ with $a_0$. Thus, in this interval of the semi-major axis, the dominant peak in the spectra of the semi-major axis is that one associated with Corotation resonance, as can be seen in the spectrum given in the blue curve at the bottom of the IPS. The initial condition for the spectrum corresponds to the initial state of Aegaeon at the date, $a_0\sim168,033.2$ km, as indicated by a vertical dashed line in the IPS, where $P_{\sigma_2}\sim1278.7$ day.

 \begin{figure}
 \centering
 \includegraphics[width=12cm]{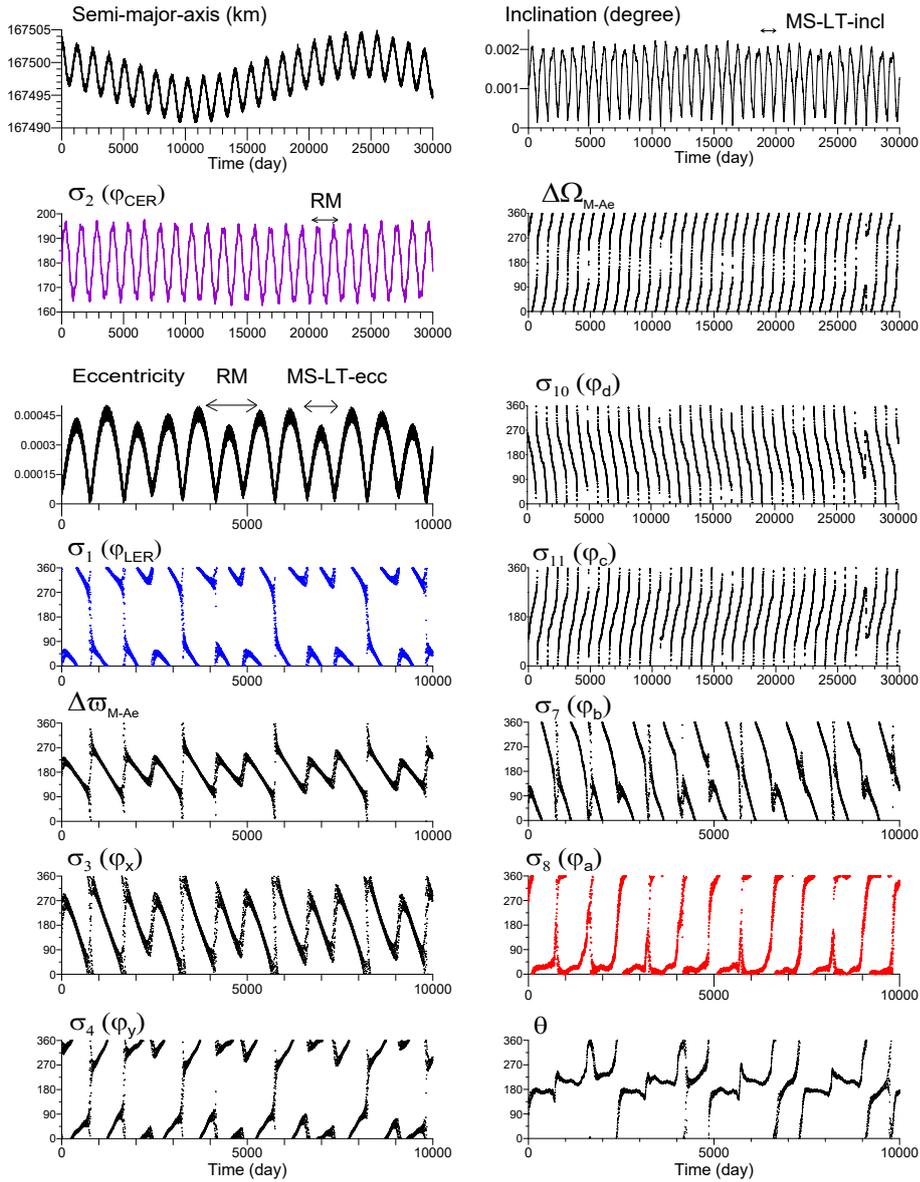}
  \caption{Time variations of geometric orbital elements and some critical arguments of the disturbing function obtained from numerical simulation of a satellite similar to Aegaeon. For sake of clarity in the interpretation of the dynamics, the x-axes of the plots are shown in two different scales depending on the plot. In parentheses, we give the notation of the angles as reported by Hedman et al. (2010). See Table 1 and Section 2 for definitions. The initial conditions have been taken from \emph{Horizons} System of ephemerides at the date 2016-01-01 and are listed in Appendix. The simulations include the Aegaeon and the mid-sized satellites up to Titan, and $J_2$ and $J_4$ of Saturn. Horizontal arrows show the main regimes of motion and have the following meaning: RM: resonant mode; MS-LT-ecc: mutual-secular long-term mode in eccentricity; MS-LT-incl: mutual-secular long-term mode in inclination.}
  \label{<F4>}
\end{figure}

The continuation of \textbf{$P_{\sigma_2}$} is interrupted at the left and right borders of the resonance, where two vertical ``barriers'' are reached at $a_0\sim168,000$ km and $a_0\sim168,070$ km. The spectra of orbits have a large number of peaks surrounding the fundamental frequencies, showing their irregular (and possibly chaotic) nature. This occurs because these regions of the phase space are located within the separatrices of the resonance (see figure 6 in Callegari and Yokoyama 2020).

The libration of corotation angle $\sigma_2$ is given in purple curve in Fig. \ref{<F4>}. Note that the time variation of the semi-major axis follows the mode of the resonance. Additionally, there is also a long-term component of $P_{42}\sim26,214.4$ day which is associated with the perturbations on the orbit of Aegaeon due to the Mimas-Tethys 4:2 mean motion resonance. Callegari et al. (2021) reported the same perturbation in the case of the satellite Methone, while Hedman et al. (2010) pointed out it in the case of Aegaeon. The 4:2 perturbation appears like an upper horizontal in IPS of the semi-major axis (Fig. \ref{<F3>}). From the spectrum of the semi-major axis, we can see that the \emph{half}-amplitude of variation of the semi-major axis due to the 4:2 resonance is comparable to that induced by corotation resonance: $\sim 3$ km. This agrees with the total amplitude of $\sim 13$ km which can be roughly estimated by inspection of Fig. \ref{<F4>}. \textbf{$H_1$}, indicated in IPS and spectra of the semi-major axis is a harmonic of $P_{42}$ with one-third of the fundamental period. (Similar result is also observed in the case of Methone Callegari et al. 2021.)

Fig. \ref{<F4>} shows the \emph{geometric} orbital elements of Aegaeon. We adopt the \emph{osculating} elements in this work, in particular, in the construction of the dynamical maps. However, the geometric elements are calculated when we analyze the individual orbits of Aegaeon and some test satellites since, in this type of problem we are dealing with in this work, osculating orbital elements may suffer large short-term variations due to $J_2$ while the corresponding geometric elements do not. This occurs in general for orbital eccentricity very close to zero, and the most affected variables are the eccentricity, mean anomaly, and argument of the pericenter. This discussion is given in detail in Callegari and Yokoyama (2020) and Callegari et al. (2021) for this sort of situation (close-in satellites revolving around a strong $J_2$ field). The geometric elements have been calculated from the vector states of the clones of Aegaeon in our simulations, applying the algorithm of Renner and Sicardy (2006).

Next, let us consider the long-term variations on the orbit of Aegaeon due to mutual-secular interactions with Mimas and secular variations due to $J_2$.

\subsection{The current orbit of Aegaeon: the mutual-secular long-term modes. }

The corotation resonance affects all variables, therefore $P_{\sigma_2}$ must also appear in the other IPSs of eccentricity and inclination, as it is indicated in Fig. \ref{<F3>}. In the case of these two latter variables, there are two additional components that are very preeminent in the spectra and are related to the following angles: $\Delta\Omega=\Omega_M-\Omega_s$ and $\Delta\varpi=\varpi_M-\varpi_s$. These variables are linked respectively to the inclination and eccentricity variables (Brouwer and Clemence 1966). Following Callegari et al. (2021), the fundamental frequencies associated with these angles are denoted by secular-mutual long-term modes (see also  Callegari and Yokoyama 2020). They rise due to the $J_2$ secular perturbations on  $\Omega_i$ and $\varpi_i$, $i$ for all satellites (and, in very small amount, due to $J_4$), and the mutual \emph{secular} and \emph{long-term} perturbations on the test satellites due to the closest mid-sized satellite, Mimas.

The loci of $P_{\Delta\varpi}$ and $P_{\Delta\Omega}$ in Fig. \ref{<F3>} are shown in the spectra of the corresponding variables. Note that the continuation of $P_{\Delta\Omega}$ and $P_{\Delta\varpi}$ are also interrupted at the chaotic borders of the resonance, but rise again beyond the separatrix since they are independent of the resonance. It is worth noting that the perturbation in the orbits of Aegaeon and therefore the principal properties of IPSs are mainly dictated by interactions with Mimas and $J_2$. Numerical experiments have shown that among all other satellites considered in simulations, Tethys has the larger contribution; in particular, its effects are significant for orbits given in the domains of the separatrices of the resonance.

In the case of Mimas-Aegaeon, the associated periods are $P_{\Delta\Omega_{M-Ae}}\sim832.20$ day and $P_{\Delta\varpi_{M-Ae}}\sim825.65$ day. See the  plots of eccentricity, inclination, $\Delta\Omega_{M-Ae}$ and $\Delta\varpi_{M-Ae}$ in Fig. \ref{<F4>}. Note that the period of ``secular'' interactions ($P_{\Delta\Omega_{M-Ae}}$ and  are $P_{\Delta\varpi_{M-Ae}}$) are smaller than the resonant period $P_{\sigma_2}\sim1278.7$ day. A similar result has been reported in the case of Enceladus-Dione 2:1 mean motion resonance in Callegari and Yokoyama (2007). The IPS of eccentricity and corresponding spectrum show also the harmonic \textbf{$H_2$} between $P_{\Delta\varpi}$ and \textbf{$P_{\sigma_2}$}, which is a linear combination of these periods such that the frequency $f_{\Delta\varpi-\sigma_2}$ has the corresponding period $P_{H_2}=P_{(\Delta\varpi-\sigma_2)}=\frac{1}{P_{\Delta\varpi}}-\frac{1}{P_{\sigma_2}}=2330.17$ day.

Figs. \ref{<F5>}a,b shown in green curves two detailed views of the geometric eccentricity and inclination given in Fig. \ref{<F4>}, where the resonant and the secular-mutual long-term modes are indicated by arrows. Pink curves are plots of analytical solutions of secular theory including Mimas-Aegaeon, $J_2$, and $J_4$ (e.g. Murray and Dermott, 1999, chapter 7). The results of linear theory show good agreement with numerical simulations of full equations of motion where several satellites are considered.

 \begin{figure}
 \centering
 \includegraphics[width=11cm]{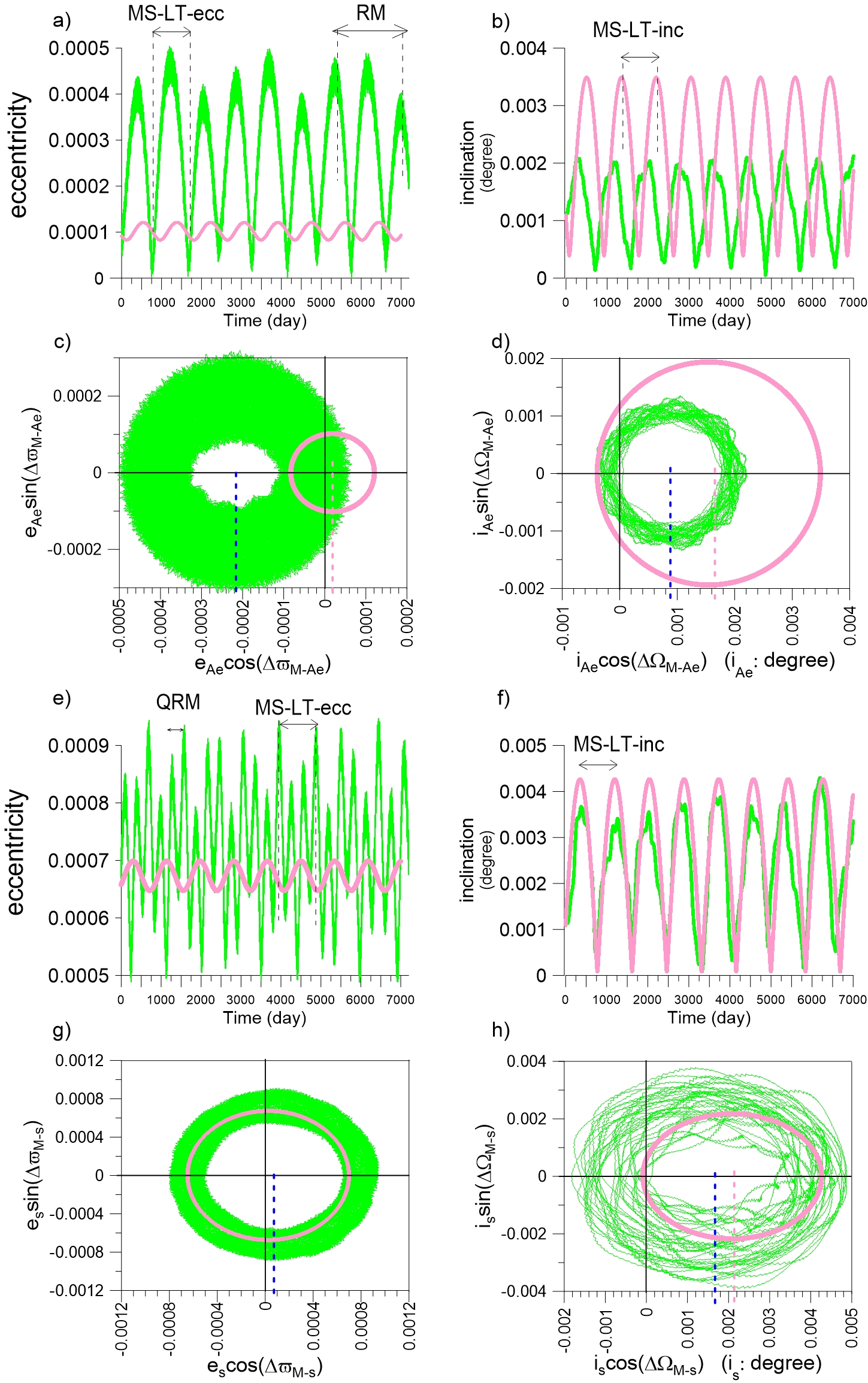}
     \caption{Green curves show geometric orbital elements of distinct orbits calculated from numerical simulations with the same model adopted in Fig. \ref{<F4>}. Pink curves result from corresponding initial conditions obtained from secular theory. The main regimes of motion are indicated by horizontal arrows (see Fig. Figs. \ref{<F4>}; QRM means quasi-resonant mode). Vertical dashed lines indicate the approximated forced centers of the orbits. \textbf{(a-d)} Current Aegaeon, starting from initial conditions at date 2016-01-01 and indicated by the magenta star in Fig. \ref{<F1>} ($a_0\sim168,033.2$ km, $e_0\sim0.0032$). Green curves in (a,b) are detailed views of the same plots given in Fig. \ref{<F4>}; \textbf{(c,d) } Projection of the orbit onto the planes ($e_{Ae}\cos(\Delta\varpi_{M-Ae}),e_{Ae}\sin(\Delta\varpi_{M-e})$) and ($i_{Ae}\cos(\Delta\Omega_{M-Ae}),i_{Ae}\sin(\Delta\Omega_{M-Ae})$). \textbf{(e-h)} The same as (a-d), in the case of an orbit outside the Corotation zone indicated by magenta triangle in  Figs. \ref{<F1>}a,b,c ($a_0=167,900$ km, $e_0=0.0025$). $s$ indicates a test satellite.}
  \label{<F5>}
\end{figure}

The green curve in Fig. \ref{<F5>}c shows the projection of the geometric orbit of Aegaeon onto the plane ($e_{Ae}\cos(\Delta\varpi_{M-Ae}),e_{Ae}\sin(\Delta\varpi_{M-e})$). Note that there is a forced component of $\sim0.00023$ (indicated by vertical dashed blue line) centered at $\Delta\varpi_{M-Ae}=\pi$ such that the projected orbit is almost tangent to the axis $e_{Ae}\cos(\Delta\varpi_{M-e})=0$. Thus, since the forced and free eccentricities have the same order the magnitude, $\Delta\varpi_{M-e}$ alternates its variation between retrograde circulation and oscillation around $\pi$, as can be seen in the plot of this angle in Fig. \ref{<F4>}. The result of the secular theory (pink curve) shows a very small forced component of $\sim0.00002$ centered at $\Delta\varpi_{M-Ae}=0$ (indicated by vertical dashed pink line in Fig. \ref{<F5>}c).

Fig. \ref{<F5>}d shows the same orbits in the plane ($i_{Ae}\cos(\Delta\Omega_{M-Ae}),i_{Ae}\sin(\Delta\Omega_{M-e})$). There is a forced component of $i_f\sim0.001$ degree (indicated by vertical dashed blue line) centered at $\Delta\Omega_{M-Ae}=0$ such that the projected orbit is almost tangent to the axis $i_{Ae}\cos(\Delta\Omega_{M-Ae})=0$. Note that the forced component in inclination is similar to the current value of the geometric inclination to the date ($i_{Ae}\sim0.001$ degree). In fact, the free inclination is very small. Though small, there are consequences of the forced mode in inclination on the orbit of Aegaeon, and further discussions will be given in Section 3.1.

The large differences between forced centers of the projected curves of the secular linear theory and the current orbit of Aegaeon in Fig. \ref{<F5>}c have an important consequence: the 7:6 Mimas-Aegaeon is responsible for the forced component in eccentricity of Aegaeon, centered at $\Delta\varpi_{M-Ae}=\pi$. Following the same arguments given in Callegari et al. (2021), we can take an orbit outside the resonance to see that now the forced mode in eccentricity due to resonance is almost null so that the projected orbit will be closer to the corresponding one provided by the secular theory. An example is shown in Figs. \ref{<F5>}e-g. The initial condition is indicated by a triangle in Fig. \ref{<F1>}. Note that the mutual-secular long-term mode is dominant, while the resonance now is replaced by a quasi-resonant state driven by a rapid oscillation (Fig. \ref{<F5>}e).

\begin{table*}
 \centering
  \caption{The arguments of the expanded disturbing function of the three-body problem up to degree four in eccentricity and inclination associated to a general ($p+1$):$p$ first-order mean motion commensurability ($p$ is an integer; see table B.4 in Murray and Dermott 1999). $p=6$ for the 7:6 Mimas-Aegaeon mean motion resonance. Symbols $i$ and $o$ refer to the inner and outer satellite in the resonant pair, respectively (such that $i$: Aegaeon, $o$: Mimas in the case of the current Mimas-Aegaeon resonance). $\varphi_x$, $\varphi_y$, $\varphi_a$, $\varphi_b$, $\varphi_c$, $\varphi_d$, $\varphi_{CER}$, $\varphi_{ILR}$ are the notations of the same corresponding angles studied in Hedman et al. (2010). $s=\sin\frac{i}{2}$, where (in this work) $i$ is the inclination of the satellite orbit with respect to the equator of Saturn, and the other symbols indicate the classical orbital elements: $\lambda$ (mean longitude), $\Omega$ (longitude of the ascending node), $\varpi$ (longitude of the pericenter), $e$ (orbital eccentricity).}
   \vspace{0.5cm}

\begin{tabular}{cccc}
\hline
Critical argument: & Argument of the cosine                     & Factors  \\
(index)            &                                            &          \\
\hline \hline

$\sigma_1$ ($\varphi_{ILR}$) &$(p+1)\lambda_o-p\lambda_i-\varpi_i$                       & $e_i$, $e_i^{3}$, $e_ie_o^{2}$ , $e_i(s_o^{2}+s_i^{2})$  \\ \hline

$\sigma_2$ ($\varphi_{CER}$) &$(p+1)\lambda_o-p\lambda_i-\varpi_o$                       & $e_o$,  $e_o^{3}$, $e_i^2e_o$, $e_o(s_o^2+s_i^2)$   \\ \hline

$\sigma_3$ ($\varphi_x$) &$(p+1)\lambda_o-p\lambda_i+\varpi_o-2\varpi_i$             & $e_i^{2}e_o$ \\ \hline

$\sigma_4$ ($\varphi_y$) &$(p+1)\lambda_o-p\lambda_i-2\varpi_o+\varpi_i$             & $e_ie_o^{2}$ \\ \hline

$\sigma_5$  &$(p+1)\lambda_o-p\lambda_i+\varpi_i-2\Omega_i$             & $e_is_i^{2}$   \\ \hline

$\sigma_6$  &$(p+1)\lambda_o-p\lambda_i+\varpi_o-2\Omega_i$             & $e_os_i^2$ \\ \hline

$\sigma_7$ ($\varphi_b$) &$(p+1)\lambda_o-p\lambda_i-\varpi_i-\Omega_o+\Omega_i$     & $e_is_is_o$  \\ \hline

$\sigma_8$ ($\varphi_a$) &$(p+1)\lambda_o-p\lambda_i-\varpi_i+\Omega_o-\Omega_i$     & $e_is_is_o$  \\ \hline

$\sigma_9$  &$(p+1)\lambda_o-p\lambda_i+\varpi_i-\Omega_o-\Omega_i$     & $e_is_is_o$   \\ \hline

$\sigma_{10}$ ($\varphi_d$) &$(p+1)\lambda_o-p\lambda_i-\varpi_o-\Omega_o+\Omega_i$    & $e_os_is_o$  \\ \hline

$\sigma_{11}$ ($\varphi_c$) &$(p+1)\lambda_o-p\lambda_i-\varpi_o+\Omega_o-\Omega_i$    & $e_os_is_o$  \\ \hline

$\sigma_{12}$ &$(p+1)\lambda_o-p\lambda_i+\varpi_o-\Omega_o-\Omega_i$    & $e_os_is_o$  \\ \hline

$\sigma_{13}$ &$(p+1)\lambda_o-p\lambda_i+\varpi_i-2\Omega_o$            & $e_is_o^{2}$   \\ \hline

$\sigma_{14}$ &$(p+1)\lambda_o-p\lambda_i+\varpi_o-2\Omega_o$            & $e_os_o^{2}$   \\ \hline

\end{tabular}
\end{table*}

\newpage

\subsection{The current orbit of Aegaeon: the Hedman's angles.}

Of particular importance among the several angles listed in Table 1 is the Lindblad angle $\sigma_1=7\lambda_M-6\lambda_{Ae}-\varpi_{Ae}$. (See Table 1; the corresponding angles with the notation adopted by Hedman et al. (2010) are also indicated at the left in (\ref{3})-(\ref{8}).). The blue curve in Fig. \ref{<F4>} shows that geometric $\sigma_1$ alternates between oscillation around zero and retrograde circulation. $\sigma_1$ can be written as a combination of the corotation angle and $\Delta\varpi_{M-e}$ in the following way. Define $\beta$ so that $\beta=7\lambda_M-6\lambda_{Ae}$. Adding and subtracting $\varpi_{M}$ in the expression of $\sigma_1$ given in Table 1 we have:
\begin{eqnarray}
\sigma_1&=&\beta-\varpi_{Ae} \nonumber\\
&=&\underbrace{\beta-\varpi_{M}}+\underbrace{\varpi_{M}-\varpi_{Ae}} \nonumber\\
&=&\hspace{0.5cm}\sigma_2 \hspace{0.5cm}+\hspace{0.5cm} \Delta\varpi_{M-Ae}. \label{2}
\end{eqnarray}
Joining with the results of Section 2.3, we conclude that since \emph{all the time} $\Delta\varpi_{M-Ae}$ oscillates around $\pi$, $\sigma_1$ oscillates around \emph{zero} because $\sigma_2$ \emph{librates} around $\pi$. When $\Delta\varpi_{M-Ae}$ circulates, the same occurs with the Lindblad angle. Inspection of the plots of the geometric angles $\Delta\varpi_{M-Ae}$, $\sigma_1$ and $\sigma_2$ in Fig. \ref{<F4>} allows us graphically to confirm this composition of fundamental modes.

Hedman et al. (2010) have already noted the above composition $\sigma_1=\sigma_2+\Delta\varpi_{M-Ae}$. Their description of the time variations of $\sigma_1$ are that Aegaeon's orbit lies close to the boundaries of the \emph{Lindblad} resonance. Here we give a more general interpretation of the problem by noting that \emph{the true and the only} resonant angle defining the 7:6 Mimas-Aegaeon resonance is the corotation angle, which \emph{always} \emph{librates} around $\pi$. $\sigma_2$, on this turn, alternates between different episodic regimes of motion due to the forced component in eccentricity in $\Delta\varpi_{M-Ae}$.

Many other geometric angles displayed in Table 1 exhibit compositions of the fundamental frequencies in their time variations similar to the case of the Lindblad angle:

\begin{eqnarray}
\varphi_x: \hspace{1cm}\sigma_{3}&=&\beta+\varpi_{M}-2\varpi_{Ae} \nonumber\\
&=&\underbrace{ \beta-\varpi_{M} } +\underbrace{(\varpi_{M}+\varpi_{M}-\varpi_{Ae}-\varpi_{Ae})} \nonumber\\
        &=&\hspace{0.4cm}\sigma_2 \hspace{0.5cm}+\hspace{0.6cm}  2\Delta\varpi_{M-Ae}, \label{3}\\
        \nonumber\\
\varphi_y: \hspace{1cm} \sigma_{4}&=&\beta-2\varpi_{M}+\varpi_{Ae} \nonumber\\
&=&\underbrace{ \beta-\varpi_{M} } \underbrace{-\varpi_{M}+\varpi_{Ae}} \nonumber\\
        &=&\hspace{0.4cm}\sigma_2 \hspace{0.5cm}-\hspace{0.2cm}  \Delta\varpi_{M-Ae}, \label{4}\\
       \nonumber\\
\varphi_b: \hspace{1cm} \sigma_{7}&=&\beta-\varpi_{Ae}-\Omega_{M}+\Omega_{Ae} \nonumber\\
        &=&\underbrace{ \beta-\varpi_{M} } + \underbrace{  \varpi_{M}-\varpi_{Ae}  }\underbrace{ -\Omega_{M}+\Omega_{Ae} }\nonumber\\
        &=&\hspace{0.5cm}\sigma_2 \hspace{0.4cm}+\hspace{0.1cm} \Delta\varpi_{M-Ae}\hspace{0.2cm}-\hspace{0.2cm}\Delta\Omega_{M-Ae},\label{5}\\
        \nonumber\\
\varphi_a: \hspace{1cm}\sigma_{8}&=&\beta-\varpi_{Ae}+\Omega_{M}-\Omega_{Ae} \nonumber\\
        &=&\underbrace{ \beta-\varpi_{M} } + \underbrace{  \varpi_{M}-\varpi_{Ae}  }+\underbrace{\Omega_{M}-\Omega_{Ae}}\nonumber\\
        &=&\hspace{0.5cm}\sigma_2 \hspace{0.4cm}+\hspace{0.1cm} \Delta\varpi_{M-Ae}\hspace{0.2cm}+\hspace{0.2cm}\Delta\Omega_{M-Ae},\label{6}\\
        \nonumber\\
\varphi_d: \hspace{1cm} \sigma_{10}&=&\underbrace{ \beta-\varpi_{M} } \underbrace{-\Omega_{M}+\Omega_{Ae}}\nonumber\\
        &=&\hspace{0.5cm}\sigma_2  \hspace{0.4cm}-\hspace{0.3cm}  \Delta\Omega_{M-Ae}, \label{7}\\
        \nonumber\\
\varphi_c: \hspace{1cm}\sigma_{11}&=&\underbrace{ \beta-\varpi_{M} } +\underbrace{\Omega_{M}-\Omega_{Ae}} \nonumber\\
        &=&\hspace{0.5cm}\sigma_2 \hspace{0.4cm}+\hspace{0.3cm} \Delta\Omega_{M-Ae}. \label{8}
\end{eqnarray}

$\sigma_4$ is the same as $\sigma_2$ with the opposite signal in the $\Delta\varpi_{M-Ae}$ component (compare (\ref{4}) and (\ref{2}) and the plots of $\sigma_1$ and $\sigma_4$ in Fig. \ref{<F4>}). $\sigma_3$ is driven mainly by $\Delta\varpi_{M-Ae}$ enhanced in amplitude by a factor 2 (compare the plots of $\Delta\varpi_{M-Ae}$ and $\sigma_3$). Note that we have added and subtracted $\varpi_{M}$ in the expression of $\sigma_3$, $\sigma_7$ and $\sigma_8$ to obtain the final forms (\ref{3}), (\ref{5}) and (\ref{6}), while in the other cases singles arrangements of their expressions (see Table 1) have been done.

$\sigma_{10}$ and $\sigma_{11}$ are driven by $\Delta\Omega_{M-Ae}$ so that they circulate in distinct directions. In general, it is difficult to note in plots of $\sigma_{10}$ and $\sigma_{11}$ the composition of $\Delta\Omega_{M-Ae}$ with the frequency of the resonance (equations (\ref{7}) and (\ref{8})), since the period of $\sigma_2$ is larger than the period of $\Delta\Omega_{M-e}$ ($P_{\sigma_2}\sim1278.75$ day, $P_{\Delta\Omega_{M-Ae}}\sim832.20$ day). However, there are some intervals of time that $\Delta\Omega_{M-e}$ \emph{oscillates around zero with small amplitude}, so that the libration of $\sigma_2$ around $\pi$ rises in the plots of $\sigma_{10}$, $\sigma_{11}$. An example of this fact can be seen at the time $\sim27,000$ day in Fig. \ref{<F4>} (compare the plots of the geometric $\sigma_{10}$, $\sigma_{11}$ and $\Delta\Omega_{M-Ae}$).

The $\sigma_8$ variation is explained by the composition of two main two modes: it is driven by $\Delta\Omega_{M-Ae}$, and shows episodic oscillations around \emph{zero}. The latter occurs due to the sum of the components of $\Delta\varpi_{M-Ae}$ and $\sigma_2$ when both are oscillating around $\pi$. Another situation occurs when $\Delta\varpi_{M-Ae}$ is not oscillating: in this case, it is circulating in the opposite direction of that $\Delta\Omega_{M-Ae}$, both with almost the same periods, $P_{\Delta\Omega_{M-Ae}}\sim832.20$ day, $P_{\Delta\varpi_{M-Ae}}\sim825.65$ day. Interestingly, their sum may also oscillate around $\pi$, as it is shown by the angle $\theta=\Delta\varpi_{M-s}+\Delta\Omega_{M-s}$ (Fig. \ref{<F4>}). The above description of $\sigma_8$ for the current orbit of Aegaeon will be clearly in Section 3.1, Fig. \ref{<F6>}, where close orbits with slightly different initial conditions will be studied.

$\sigma_7$ is driven by $\Delta\Omega_{M-Ae}$ component in opposite direction of that $\sigma_8$ (compare equations (\ref{5}) and (\ref{6})). In the case of simultaneous oscillations of $\Delta\varpi_{M-Ae}$ and $\sigma_2$ the descriptions given above for $\sigma_8$ are valid, but the same is not true in the case when $\Delta\varpi_{M-Ae}$ circulates, since due to changed sign in $\Delta\Omega_{M-Ae}$, their components are superposed.

 \begin{figure}
\centering
   \includegraphics[width=11cm]{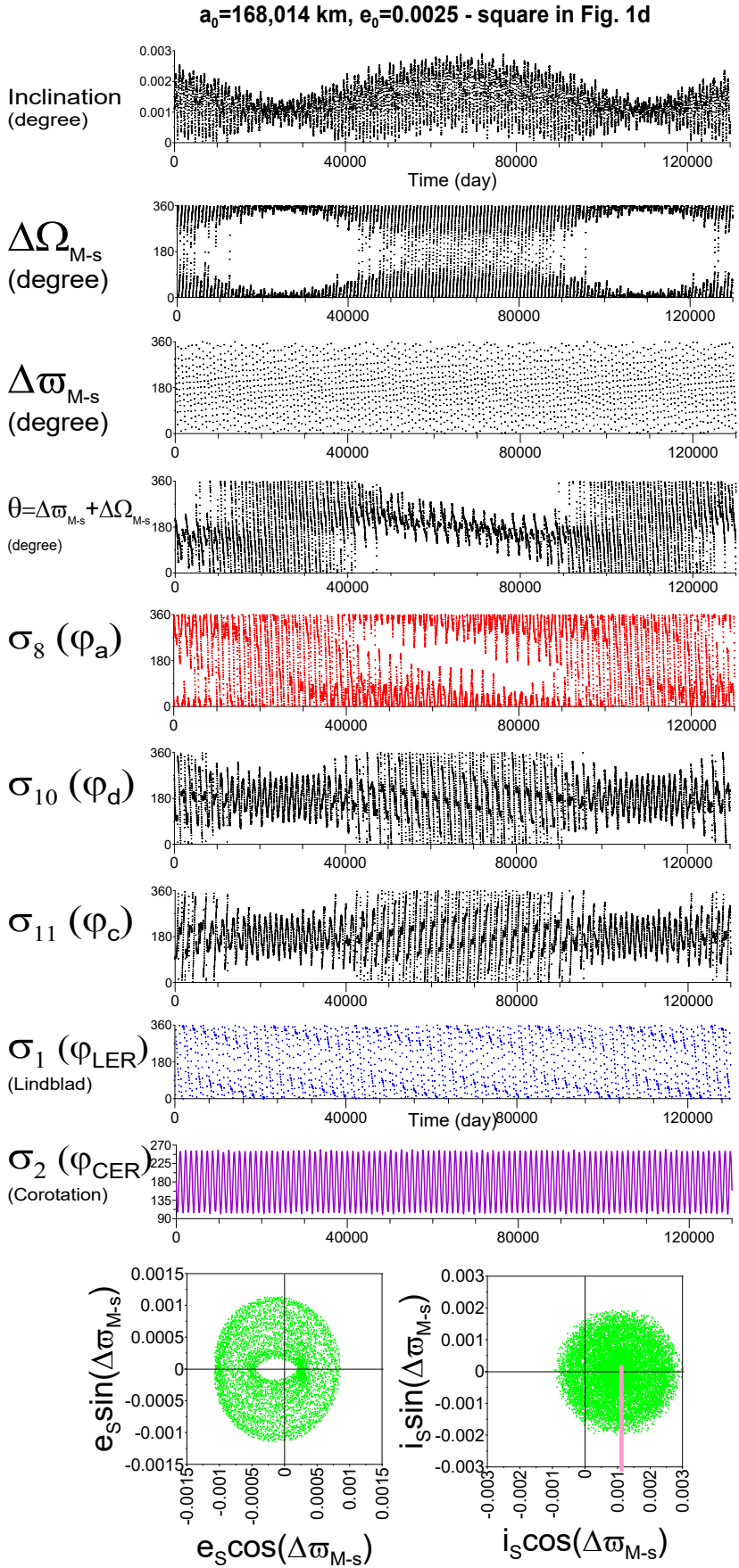}
     \caption{Critical angles and geometric elements for an orbit with $e_0=0.0025$ and $a_0=168,014$ km - magenta square in Fig.\ref{<F1>}. Bottom green curves: projected orbit onto the planes ($i_s\cos(\Delta\Omega_{M-s}),i_s\sin(\Delta\Omega_{M-s})$ (right) and ($e_s\cos(\Delta\varpi_{M-s}),e_s\sin(\Delta\varpi_{M-s})$ (left), where $s$ is for a test satellite.}
     \label{<F6>}
\end{figure}

 \begin{figure}
\centering
   \includegraphics[width=11cm]{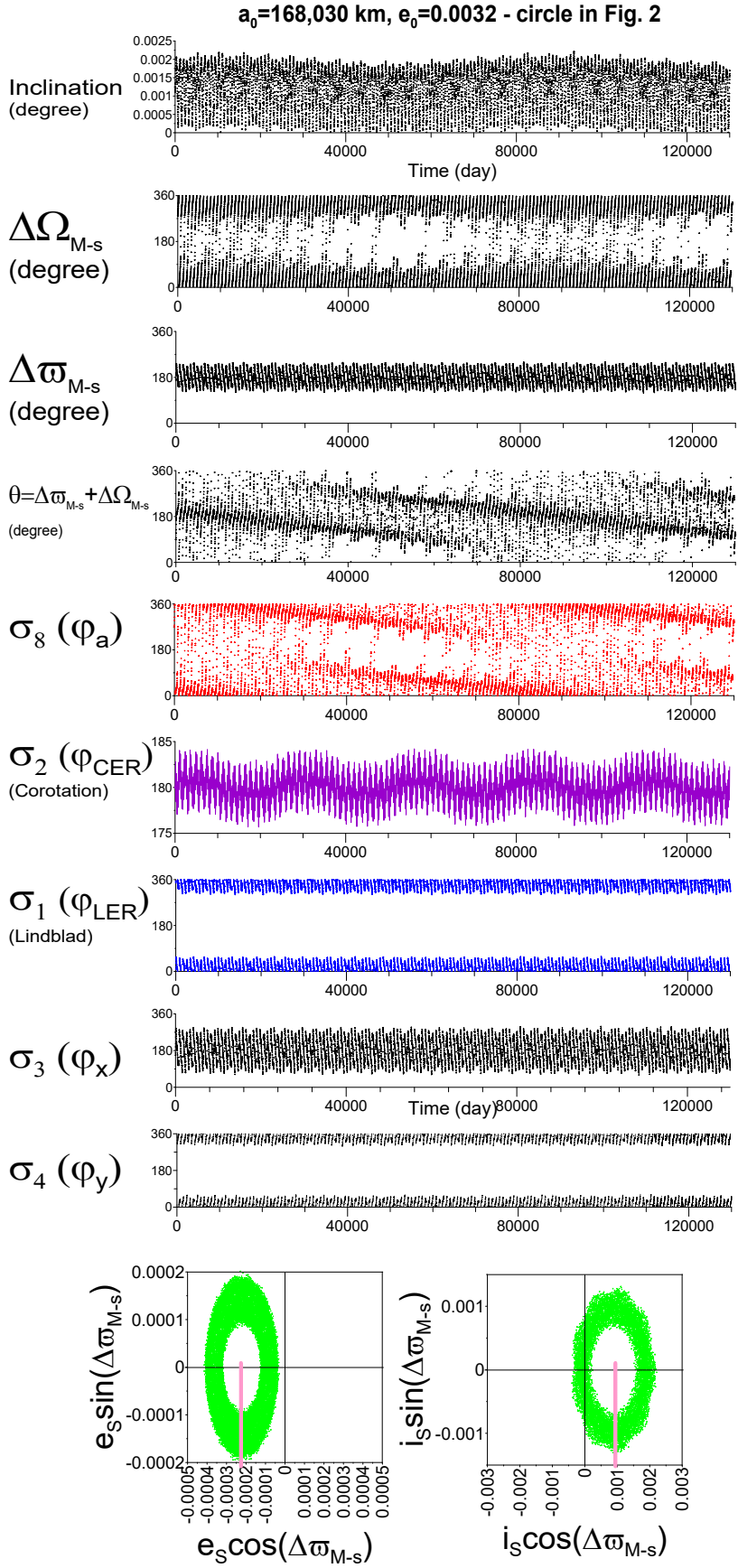}
     \caption{Critical angles and geometric elements for an orbit with $e_0=0.0032$ and $a_0=168,030$ km - open circle in Fig. \ref{<F1>}. All definitions are given in Table 1, Figures \ref{<F1>} and \ref{<F6>}. }
     \label{<F7>}
\end{figure}

\section{Additional analysis within and outside the Corotation zone}

\subsection{The $\texttt{D}$ dark strip}

In this section, let us investigate the nature of the dark structure $\texttt{D}$ located inside the Corotation zone and very close to the orbit of Aegaeon. As pointed out before, this black strip inside the Corotation zone is only present in the dynamical maps constructed with Fourier analyses of the orbital inclination (compare Figs. \ref{<F1>}a,b and Fig. \ref{<F1>}c). In the following, we will give examples that show that the $\texttt{D}$ region is associated with induced time variations in inclination due to a forced component in this orbital element (e.g. Fig. \ref{<F5>}d).

Fig. \ref{<F6>} displays results of an orbit with $e_{0}=0.0025$ and $a_{0}=168,014$ km. This position is indicated by the magenta square in Fig. \ref{<F1>}d, so we are choosing an initial condition over the dark strip. The two first plots at the top in Fig. \ref{<F6>} show that the time variation of the inclination is linked to $\Delta\Omega_{M-s}$ in such a way that when the latter oscillates around zero the former has the minimum variation ($s$ for a test satellite). Green points plotted at the bottom-right in Fig. \ref{<F6>} are the projection of the orbit onto the plane ($i_s\cos(\Delta\Omega_{M-s}),i_s\sin(\Delta\Omega_{M-s})$. There is a cloud of points concentrated around the forced inclination $i_f\sim0.001$ degree. Contrary to the current orbit (green curve in Fig. \ref{<F5>}d), where only occasionally $\Delta\Omega_{M-Ae}$ oscillates, now a much more evident forcing in inclination is seen. Thus, the oscillation of $\Delta\Omega_{M-s}$ is related to the forced component in inclination.

The consequences of the forced component in inclination on the time variations of $\sigma_{10}$, $\sigma_{11}$, and $\sigma_8$ are visible and easier explained than before (Section 2.3). In the cases of $\sigma_{10}$, $\sigma_{11}$, all the time $\Delta\Omega_{M-s}$ oscillates around zero, $\sigma_2$ resonant libration component around $\pi$ rises (recall equations (\ref{7}) and (\ref{8})).

Let us to consider $\sigma_8=\sigma_2+\Delta\varpi_{M-s}+\Delta\Omega_{M-s}$ (equation (\ref{5})). We have that $\Delta\varpi_{M-s}$ always circulates in this case, due to larger free eccentricity, despite forced eccentricity (see projected orbit at bottom-left in Fig. \ref{<F6>}. Thus, two regimes of variation of $\sigma_8$ can occur depending on the behavior of $\Delta\Omega_{M-s}$: i) when $\Delta\Omega_{M-s}$ oscillates around zero, $\sigma_8$ circulates, driven by the $\Delta\varpi_{M-s}$ circulation mode; ii) when $\Delta\Omega_{M-s}$ circulates, its sum with $\Delta\varpi_{M-s}$ oscillates around $\pi$, as indicated by $\theta=\Delta\varpi_{M-s}+\Delta\Omega_{M-s}$ in Fig. \ref{<F6>}. Thus, by adding $\sigma_2$, which librates around $\pi$, $\sigma_8$ oscillates around \emph{zero}. Inspection of the orbit over $\texttt{D}$ strip with initial condition indicated by magenta crux in Fig. \ref{<F1>} shows the same main results reported above so that there is a symmetry with respect to line of the equilibrium solution associated to corotation \textbf{C}.


The elements shown in Fig. \ref{<F7>} correspond to the open circle in Fig. \ref{<F1>}d. As expected, the amplitude of libration of the corotation angle is very small since the initial condition is located almost inside the corotation center, \textbf{C}, which gives the loci of the equilibrium of the resonance. The green curve at the bottom in Fig. \ref{<F7>} shows a strong forced component in eccentricity so that $\Delta\varpi_{M-s}$ oscillates around $\pi$. Now three other critical angles oscillate around different centers: $\sigma_1$ ($\varphi_{LER}$), $\sigma_3$ ($\varphi_x$), $\sigma_4$ ($\varphi_y$). This example resembles the case of the current orbit of the satellite Methone, as studied and fully explored in Callegari et al. (2021). $\sigma_1$, $\sigma_3$, $\sigma_4$ are explained by inspection of (\ref{2}), (\ref{3}), (\ref{4}), respectively, by noting that $\Delta\varpi_{M-s}$ oscillates around $\pi$ in this case. It is important to note that this initial condition differs by only $\sim3$ km in the semi-major axis compared to the current orbit of Aegaeon (magenta star in Fig. 2). Recall the complex  orbit of Aegaeon studied in Section 2.4.

\begin{figure}
\centering
\includegraphics[width=.7\columnwidth,angle=270]{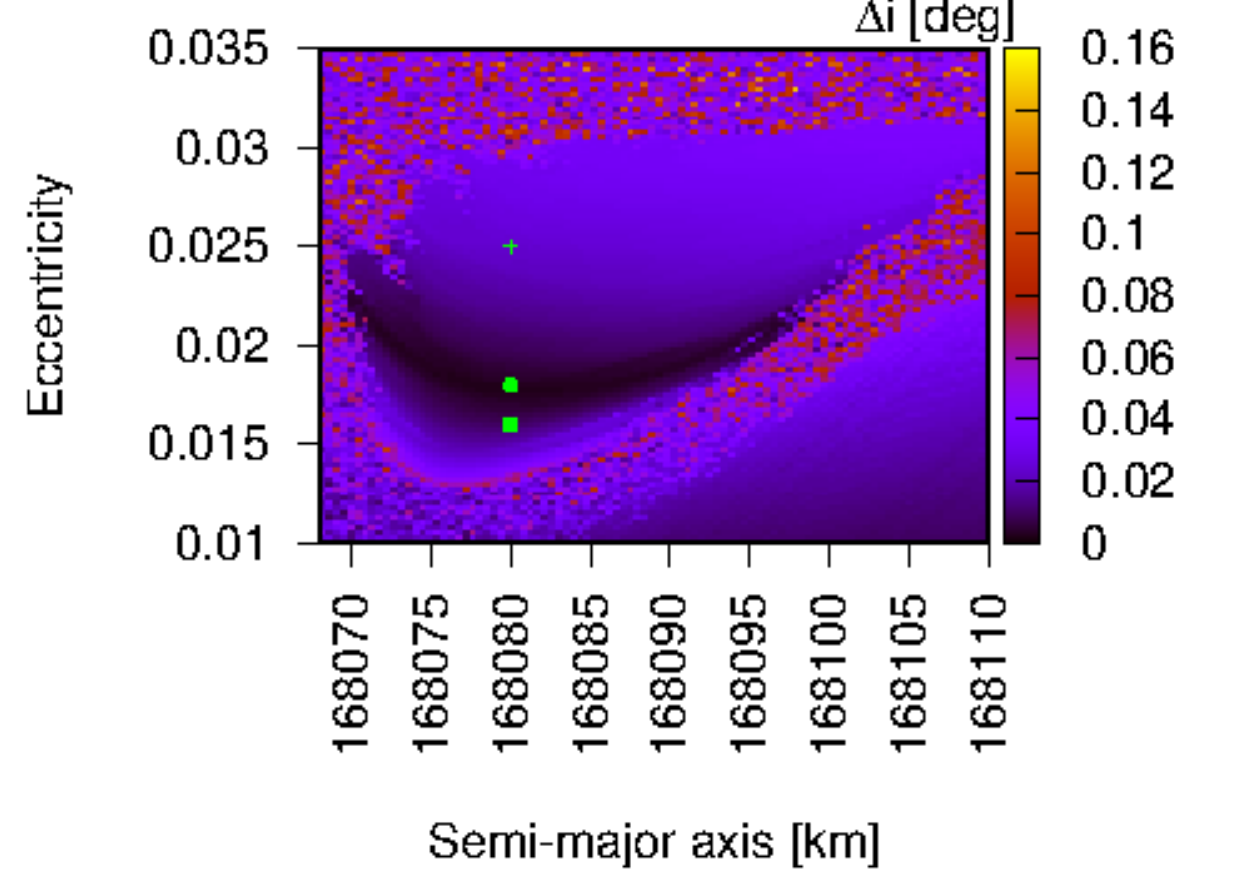}
\caption{\footnotesize{A detailed view of the region $S$ indicated in Fig. \ref{<F1>}c, where now we calculate the maximum variation in orbital inclination ($\Delta i$) for each initial condition in the map. The green symbols represent the adopted initial values for individual runs (see text for details).}}
\label{mapa2}
\end{figure}

\begin{figure}
\begin{center}
\includegraphics[width=0.65\columnwidth,angle=270]{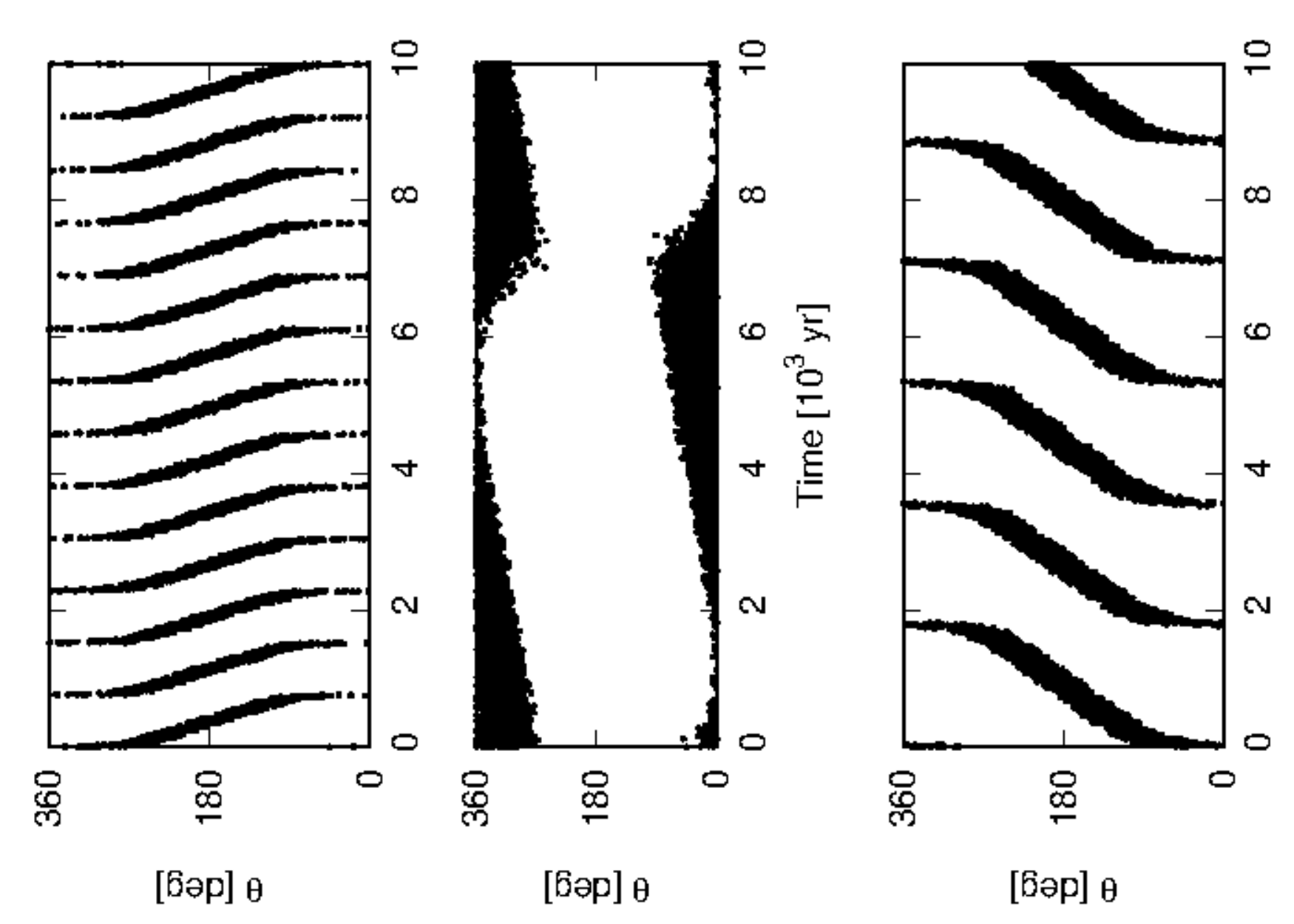}
\includegraphics[width=0.65\columnwidth,angle=270]{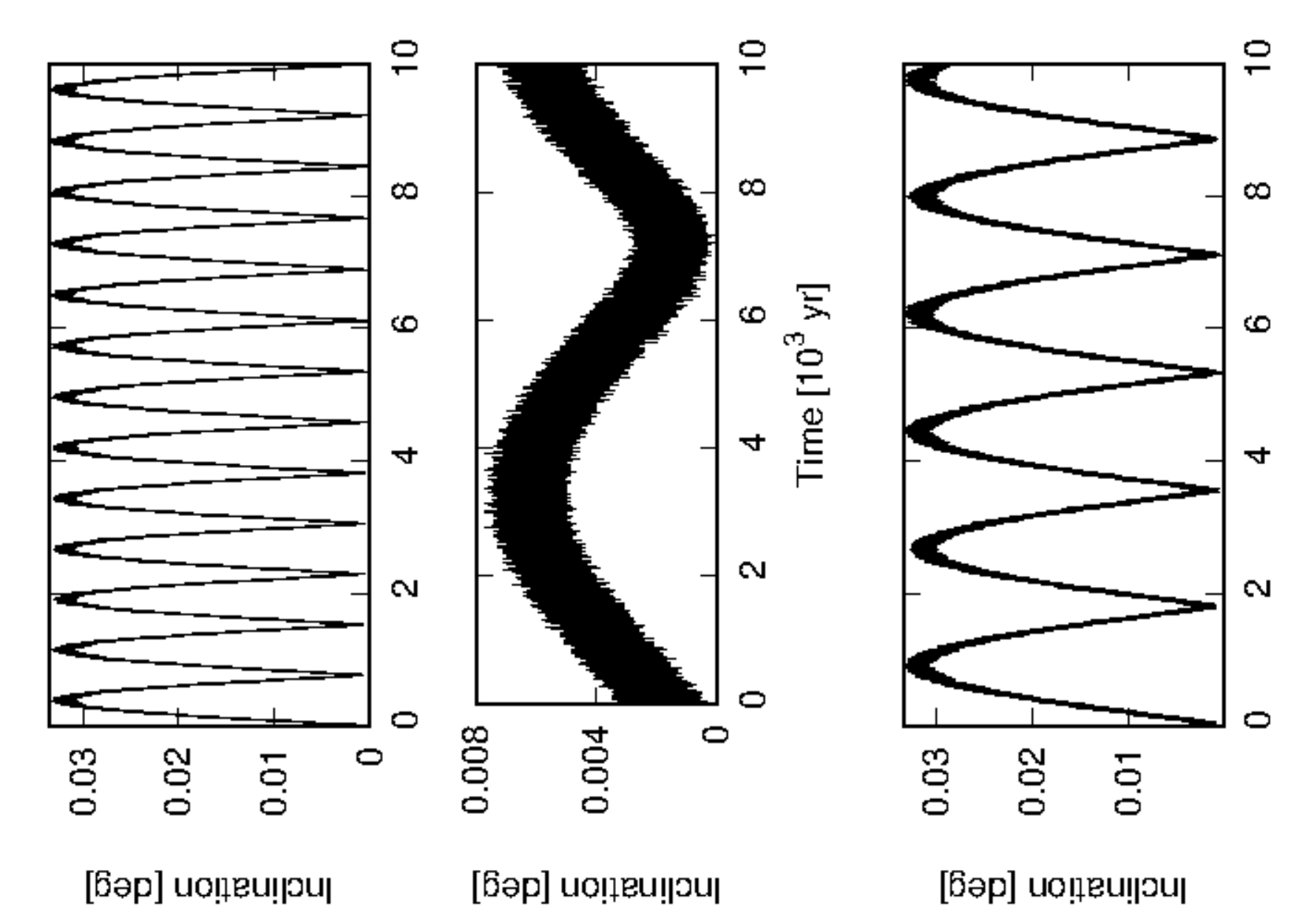}
\includegraphics[width=0.65\columnwidth,angle=270]{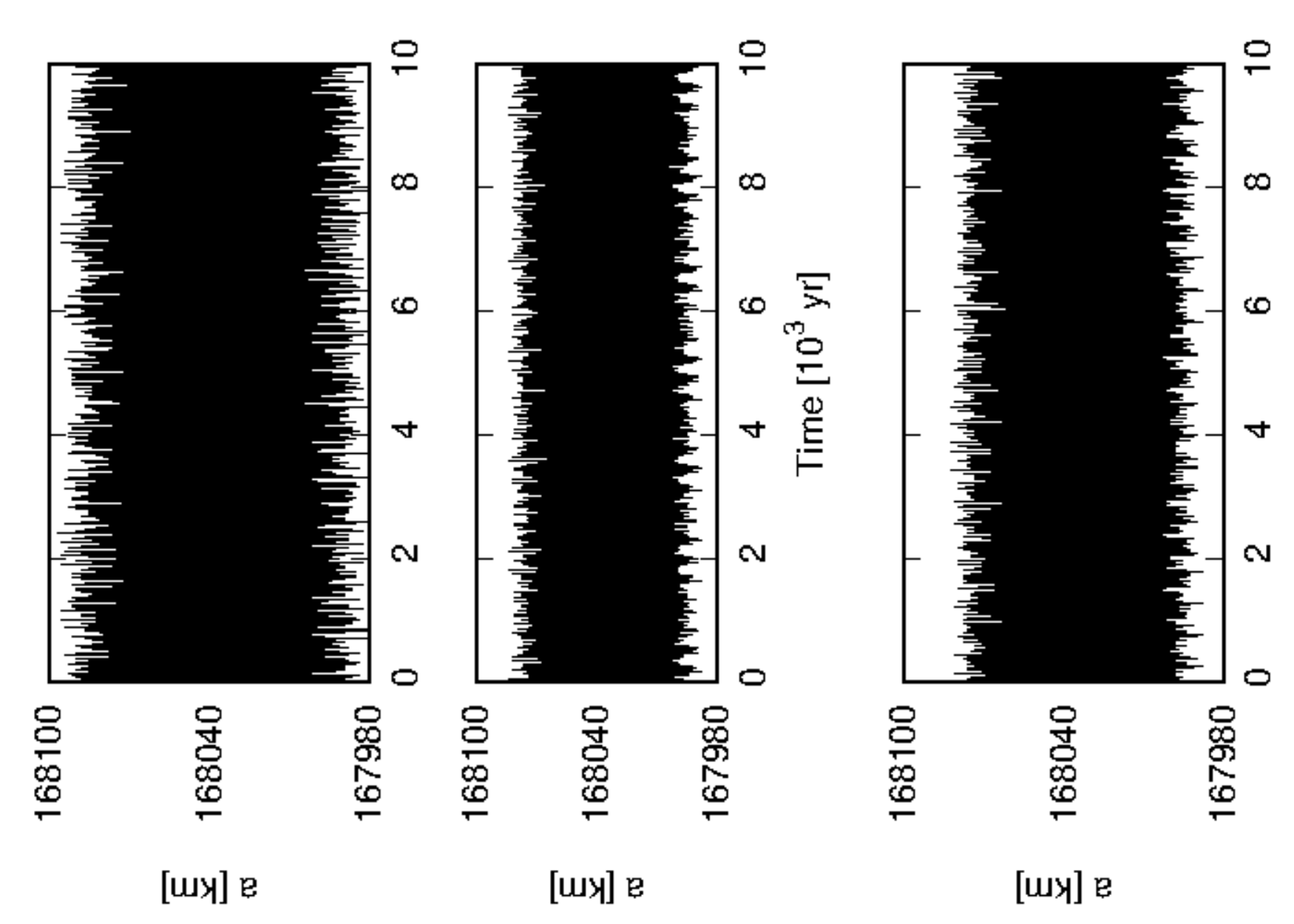}
\includegraphics[width=0.65\columnwidth,angle=270]{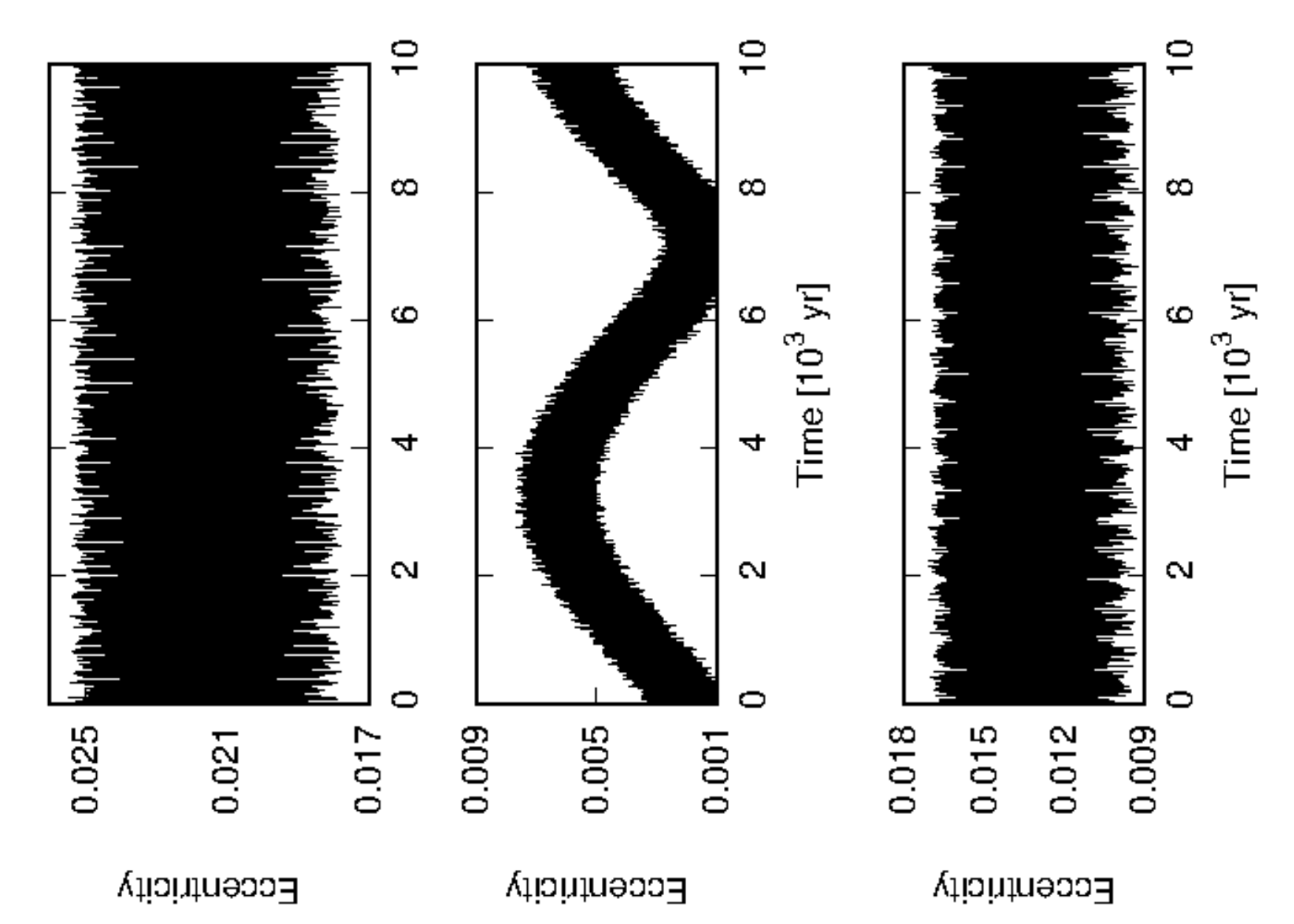}
\caption{\footnotesize{(\textit{Top-Left}): Long-term variation of $\theta=\Delta\varpi_{M-s}+\Delta\Omega_{M-s}$ for three individual numerical simulations. From top to bottom: crux, circle, and square symbols in Fig. \ref{mapa2}. (\textit{Top-Right}): The same now for the long-term variation of the orbital inclination. (\textit{Bottom-Left}): Semi-major axis. (\textit{Bottom-Right}): Orbital eccentricity.} Note the different scales of the y-axes of the plots of eccentricity and inclination.}
\label{theta-incli}
\end{center}
\end{figure}

\subsection{The \textbf{S} region}

As pointed out in Fig. \ref{<F1>}c (Section 2.1), there is a large region of the phase space at the top of the Corotation zone where regular motion occurs. Fig. \ref{mapa2} shows a new dynamical map close to the occurrence of the \textbf{S} region. We performed $10,000$ numerical simulations considering the perturbation of Mimas, Enceladus, Tethys, and Dione, plus Saturn's oblateness including $J_2$ and $J_4$ coefficients. To investigate the variation of the orbital elements within this region, we choose three initial conditions such that $a=168,080$ km and $e=0.016, 0.018, 0.025$, keeping other orbital elements equal to the Aegaeon's ones at 2016/01/01. These values correspond to eccentricities at the bottom (square), top (crux) and inside (circle) the darkest zone (see green symbols in Fig. \ref{mapa2}). Each initial condition was integrated for 200 yr, representing 77,513 Mimas orbital periods.


Fig. \ref{theta-incli} shows the time variation of the angle $\theta=\Delta\varpi_{M-s}+\Delta\Omega_{M-s}$ and other orbital elements for the above-mentioned initial conditions. $\theta$ is the same angle explored in Section 2.4. On one hand, the simulation starting with the highest (lowest) value of eccentricity indicates a retrograde (prograde) circulation of $\theta$. On the other hand, for $e=0.018$, $\theta$ oscillates around zero with amplitude $\sim100^{\circ}$ (note the difference between the integration times). It is important to note that the orbital inclination oscillates with the same period of $\theta$. Moreover, the period of oscillation strongly depends on the initial value of the eccentricity.


A numerical inspection of a large set of initial conditions within the investigated domain allowed us to show that, for those initial positions in the map at the top (bottom) of the darkest region, the angle $\theta$ has a retrograde (prograde) circulation. This indicates that the dark structure inside the \textbf{S} region is associated with a kind of separatrix linked to the variation of $\theta$.

The time variations of the semi-major axis and eccentricity do not show any special behavior within the region highlighted in Fig. \ref{mapa2}. In particular, these both variables are not confined (see Fig. \ref{theta-incli}) in the volume of phase space encompassing region \textbf{S}, ruling out therefore the possibility of the existence of any high-order mean motion resonance.

Further investigations of the dynamics inside this \textbf{S} region are outside of our initial goals in this work. In conclusion, it is a region related to a secular motion of the system, where the variable $\theta=\Delta\varpi_{M-s}+\Delta\Omega_{M-s}$ can be almost stationary oscillating zero, suffering also transition of circulation regimes around the loci of the dark strip.

\section{Conclusions and Discussions}

Motivated by the richness of the dynamic environment of Aegaeon first reported in Hedman et al. (2010), in this work we study the current orbit of this small moon with extended data and techniques of a non-linear system. We performed numerical integrations of thousands of clones of the satellite in the close vicinity of the current orbit, allowing us to obtain an in-deep study of the phase space of the 7:6 Mimas-Aegaeon resonance. This resonance is well-characterized by the libration of the corotation angle $\sigma_2$ around $\pi$, what means that the conjunctions between Mimas and Aegaeon occur in a line which always oscillates around the longitude of the apocenter of Mimas.

We explain the anomalous transitions between circulation and oscillation of the Lindblad angle $\sigma_1=7\lambda_M-6\lambda_{Ae}-\varpi_{Ae}$ ($\varphi_{LER}$ in Hedman et al. 2010), and the angles $\sigma_3$ ($\varphi_x$), $\sigma_4$ ($\varphi_y$), $\sigma_7$ ($\varphi_b$), $\sigma_8$ ($\varphi_a$), displayed in Fig. \ref{<F4>}. We have shown that the alternations of regimes of motion of these angles are consequences of a forced component in eccentricity in the orbit of Aegaeon due to resonance, leading to the angle $\Delta\varpi_{M-Ae}=\varpi_M-\varpi_{Ae}$ oscillates episodically around $\pi$. The forced eccentricity is $e_f\sim0.0002$, and has a similar magnitude to the free eccentricity, as conjectured in Hedman et al. (2010).

In the cases of $\sigma_{11}$ ($\varphi_c$) and $\sigma_{10}$ ($\varphi_d$), the relative motion of the longitudes of the ascending nodes of Mimas and Aegaeon, $\Delta\Omega_{M-Ae}=\Omega_M-\Omega_{Ae}$, plays an important role (Hedman et al. 2010). In particular, we have shown that this synodic angle $\Delta\Omega_{M-Ae}$ may episodically oscillate around zero. We explain this anomalous variation of relative ascending nodes by determining a forced component in inclination of the orbit of Aegaeon $i_f\sim0.001$ degree, a value similar to the current geometric inclination to the date so that the free inclination is very small. It is worth noting that the forced value of the orbital inclination of Aegaeon is due to long-term perturbations with Mimas, not the 7:6 resonance.

Our interpretations of the perturbations of the orbit of Aegaeon are based on the fundamental frequencies which, after being determined numerically with spectral analysis, can be physically explained in terms of resonant, secular, and long-term perturbations mainly due to Mimas and $J_2$. It is not required the co-existence of multiple resonances acting in the orbit of Aegaeon, avoiding in this case potential signals of chaotic motion in its current orbit. Moreover, we have shown that the main perturbations are due to Mimas, the term $J_2$ of the Saturn gravitational field, and with weaker contribution, Tethys.

Similar results have been found and discussed in the cases of Anthe and Methone, which also have very regular orbits trapped within their respective corotation resonances with Mimas (see Callegari and Yokoyama 2020, Callegari et al. 2021, and references therein). However, due to the complexity of the resonant perturbations acting on the orbit of Aegaeon, at first glance, this would make the 7:6 resonance with Mimas distinct from Anthe's and Methone's, but it is not the case. In fact, the Corotation zone is isolated in the phase space, reaching eccentricities $\sim0.025$ and semi-major axis given in the interval $\sim 168,000-168,060$ km, as we could determine in the numerical mapping of the phase space. The only main difference which distinguishes the case of Mimas-Aegaeon is the forced inclination. Individual analyses of orbits at the region the $\texttt{D}$ (Figs. \ref{<F1>}a,b) show that ``instabilities'' mapped at the bottom part of the Corotation zone are related to this forced component in inclination of test satellites. The maps shown in Fig. \ref{<F1>}b,c confirm this, where the analyzed variables were the semi-major axis and eccentricity, and the dark region is not present.

A key characteristic of the Mimas-Aegaeon resonance when compared with the other two corotation resonances is the closer proximity of the Aegaeon's orbit to the center of the 7:6 resonance, a result highlighted in Hedman et al. (2010) and confirmed in our numerical mappings of the loci of the periodic orbit associated to the corotation resonance (see Fig. \ref{<F1>}d). Hedman et al. (2010) conjectured that interactions of Aegaeon with other objects in that region would explain its current orbital configuration (Aegaeon is immersed in an arc within the G-ring). 

Far from the current orbit of Aegaeon and its neighborhood, we have found the existence of a regime of regular motion surrounding the separatrices of the 7:6 Corotation zone. A similar regime of motion has been already been in Callegari and Yokoyama (2010a) in the case of the satellite Pallene. No mean motion resonance is associated with those sites, and inspection of individual orbits in the interior of the \textbf{S} region reveals an equilibrium configuration of the secular angle $\theta=\Delta\varpi_{M-s}+\Delta\Omega_{M-s}$ ($s$ for test satellite), which is linked to the inclination of the orbital plane.

\begin{acknowledgements}
Thanks to XX Brazilian Colloquium on Orbital Dynamics (2021 virtual Edition), and Tadashi Yokoyama.
\end{acknowledgements}







\section*{Appendix: Initial conditions and parameters}

Table 2 gives the initial osculating elements and masses of the mid-sized satellites of Saturn and Aegaeon provided by \emph{Horizons} system of ephemerides at date January 01, 2016.

The physical data for Saturn are $M_S=5.6834\times10^{26}$ kg, $R_S=60,268\pm4$ km (equatorial radius), $J_2=0.01629071$ and $J_4=-0.0009358$ (Jacobson et al. 2006).
\begin{table*}[h]
 \centering
\caption{Orbital elements and masses of Saturnian satellites taken from the \emph{Horizons} system refereed to date 2016-January-01, 00:00. Values collected from the update August 08, 2019. For Mimas and Aegaeon, we also show the geometric elements calculated as described in Section 2. \emph{The assumption of significant digits is purely formal.} $\omega$, $\Omega$ are the argument of the pericenter and the longitude of ascending node, respectively.}
\vspace{0.5cm}
 \begin{tabular}{ccccc}
Satellite& semi-major axis  (km)            &  eccentricity                & inclination (degree)       &    mass (kg) \\
         &  $\omega$ (degree)                &    $\Omega$ (degree)         & mean anomaly (degree)               &                     \\
\hline
\hline
Mimas         &  186,025.6453660664&0.02002833386117072&1.567961379446476 &      3.75    \\
(osculating)  &  149.7709076854479 &92.64771224778193  &72.91866912551491 &              \\
\\
Mimas         &  185,547.514530160&0.01951689931353337&1.56488701023253   &          \\
(geometric)   &  142.620036727989  &92.5697135272668      &80.1385659447950 &         \\

\hline
Aegaeon       &  168,033.225440921 &0.003212213566664084&  0.001095303714535553            &\\
(osculating)  &  305.3055865889405 &172.1125168939375  & 1.283184314678524      &     \\
\\
Aegaeon       &  167,502.954276314        &0.00009142792409366025   &0.001094060176922162   &$m^{(*)}$\\
(geometric)   &  254.142530026103 &172.200360310354     & 52.3583504734714       &     \\

\hline
Enceladus    &     238,410.9497573420&0.005409700624203072 & 0.06300678158019136&10.805 \\
             &  254.6089765023341&107.0935554299123&54.19418982449177   &      \\
\hline
Tethys       &   294,975.1415797813 & 0.001057322770416529 & 1.093415017063881& 61.76  \\
             & 155.4378178629212 & 183.7764009032868&353.8714568422450  &\\
\hline
Dione        &   377,651.4603631170 &0.001723860642064985 &0.02979448678882671  &109.572\\
             &  161.8670464404854  &171.2052945850907 & 228.2313154582474 &\\
\hline
Rhea        &   5.272137290588300&0.01012735341245330 &0.3557125312486978  &230.9  \\
            &  18.62640796053094& 194.5283445769434 &27.80026501495894&    \\
\hline
Titan       &   1221,961.119150759  &0.02868590608459553&0.4023490211172993 & 13455.3\\
            &324.3272061447084  &247.8889323568444 &323.9021813775230              &\\
\hline
\end{tabular}

$(^{*})$ Not provided in \emph{Horizons} system. $m\sim7.9\times10^{10}$ kg (Rodr\'{i}guez and Callegari 2021).
\end{table*}


\begin{thebibliography}{}


Brouwer, D. and Clemence, G.M.. Methods of Celestial Mechanics (Academic Press, New York) (1961).\\
\\




Callegari Jr., N.,  Yokoyama, T.. Dynamics of Two Satellites in the 2:1 Mean-Motion Resonance: application to the case of Enceladus and Dione. Celest. Mech. Dyn. Astr. \textbf{98}, 5-30 (2007).\\
\\




Callegari Jr., N.,  Yokoyama, T.. Long-term dynamics of Methone, Anthe and Pallene. Icy Bodies of the Solar System, Proceedings of the International Astronomical Union, IAU Symposium, \textbf{263}, 161-166 (2010a).\\
\\

Callegari Jr., N.,  Yokoyama, T.. Numerical exploration of resonant dynamics in the system of Saturnian inner Satellites. Planetary and Space Science \textbf{58}, 1906-1921 (2010b).\\
\\

Callegari Jr., N.,  Yokoyama, T.. Dynamics of the 11:10 Corotation and Lindblad Resonances with Mimas, and Application to Anthe, Icarus, \textbf{348}, 113820 (2020).\\
\\

Callegari Jr., N., Rodr\'{i}guez, A., Ceccatto, D. T.. The current orbit of Methone (S/2004 S 1).
Celest. Mech. Dyn. Astr. \textbf{133:49} (2021).\\
\\

\\
Charnoz, S.; Salmon, J.; Crida, A.. The recent formation of Saturn's moonlets from viscous spreading of the main rings. Nature, Volume 465, Issue 7299, pp. 752-754 (2010).\\
\\




El Moutamid, M.; Renner, S.; Sicardy, B.. Coupling between corotation and Lindblad resonances in the elliptic planar three-body problem.
Celest Mech Dyn Astr, \textbf{118}, 235-252 (2014).\\
\\

Everhart, E.. An efficient integrator that uses Gauss-Radau spacings. In: IAU Coloquium {\bf 83}, 185-202 (1985).\\
\\
\\
Hedman, M. M., Murray, C. D., Cooper, N. J., Tiscareno, M. S., Beurle, K., Evans, M. W., Burns, J. A.. Three tenous rings/arcs for three tiny moons. Icarus \textbf{199}, 378-386 (2009).\\
\\

Hedman, M. M., Cooper, N. J., Murray, C. D., Beurle, K., Evans, M. W., Tiscareno, M. S.,  Burns, J. A.. Aegaeon (Saturn LIII), a G-ring object. Icarus \textbf{207}, 433-447 (2010).\\
\\

Jacobson, R. A., Antreasian, P. G., Bordi, J. J., Criddle, K. E., Ionasescu, R., Jones, J. B., Mackenzie, R. A., Meek, M. C., Parcher, D., Pelletier, F. J., Owen Jr., W. M., Roth, D. C., Roundhill, I. M., Stauch, J. R.. The Gravity Field of the Saturnian System from satellite observations and spacecraft tracking data. The Astronomical Journal \textbf{132}, 2520-2526 (2006).\\
\\

Madeira, G., Sfair, R., Mour\~{a}o, D. C., Giuliatti Winter, S.M.: Production and fate of the G ring arc particles due to Aegaeon (Saturn LIII). Mon. Notices Royal Astron. 475(4), 5474–5479 (2018).\\
\\

Mun\~{o}z-Guti\'errez, M. A.; Giuliatti Winter, S.. Long-term evolution and stability of Saturnian small satellites: Aegaeon, Methone, Anthe and Pallene.
Monthly Notices of the Royal Astronomical Society \textbf{470}, 3750-3764 (2017).\\
\\

Murray, C. D, Dermott, S. F.. Solar System Dynamics, Cambridge University Press (1999).\\
\\

Porco, C. C.. S/2004 S 1 and S/2004 S 2. IAU Circ. 8401 (2004 August 16) (2004).\\
\\

Porco, C. C.. S/2007 S 4. IAU Circ. 8857 (2007 July 18) (2007).\\
\\

Porco, C. C.. S/2008 S 1. IAU Circ. 9023 (2009 March 3) (2009).\\
\\

Porco, C. C., Thomas, P. C., Weiss, J. W., Richardson, D. C.. Saturn's Small Inner Satellites: Clues to Their Origins. Science \textbf{318}, 1602-1607 (2007).\\
\\


Press, W. H., Teukolsky, S. A., Vetterling, W. T., B. P. Flannery. Numerical Recipes in Fortran 77. Cambridge University Press (1996).\\
\\


Renner, S., Sicardy, B.. Use of the Geometric Elements in Numerical Simulations. Celestial Mechanics and Dynamical Astronomy \textbf{94}, 237-248 (2006).\\
\\

Rodr\'{i}guez, A., Callegari Jr., N.. Dynamical stability in the vicinity of Saturnian small moons.
The cases of Aegaeon, Methone, Anthe and Pallene. Monthly Notices of the Royal Astronomical Society \textbf{506}, 5093-5107 (2021).\\
\\

Rodr\'{i}guez, A., Callegari Jr., N., Gimenez, K.. The migration of Mimas and the implications for the resonant motion of small Saturnian moons. Celest. Mech. Dyn. Astr. submitted - this Issue (2022).\\
\\
Spitale, J. N., Jacobson,  R. A., Porco, C. C., Owen, Jr, W. M.. The Orbits of Saturn's Small Satellites Derived from Combined Historic and Cassini Imaging Observations. The Astronomical Journal \textbf{132}, 792-810 (2006).\\
\\


Thomas, P. C., Helfenstein, P.. The small inner satellites of Saturn: Shapes, structures and some implications. Icarus \textbf{344}, 113355 (2020).\\
\\
\end{thebibliography}
\end{document}